\documentclass[aps,eqsecnum,prd,twocolumn]{revtex4}
\usepackage{graphics,graphicx}
\usepackage{amsmath}
\usepackage{amssymb,latexsym,mathrsfs}
\usepackage{hyperref}

\def\bea{\begin{eqnarray}}
\def\eea{\end{eqnarray}}
\def\ba{\begin{array}}
\def\ea{\end{array}}

\def\beq{\begin{equation}}
\def\eeq{\end{equation}}

\begin{document}

\title{Comparison of decoherence and Zeno dynamics from the context of weak measurement for a two level atom traversing through squeezed vacuum}

\author{Samyadeb Bhattacharya$^{1}$ \footnote{sbh.phys@gmail.com}}
\affiliation{$^{1}$Physics and Applied Mathematics Unit, Indian Statistical Institute, 203 B.T. Road, Kolkata 700 108, India \\}

\vspace{2cm}
\begin{abstract}

\vspace{1cm}

\noindent Decay parameter of coherence and population inversion are calculated from the master equation of a two level atom tunneling through a squeezed vacuum. Using those parameters, the timescales for decoherence and zeno effect are calculated in the weak measurement scheme. By comparing those timescales, a certain condition has been found for sustainable coherent dynamics.

\vspace{2cm}


\vspace{1cm}

\end{abstract}

\vspace{1cm}

\maketitle

\section{Introduction}

The ability to preserve coherence in a quantum mechanical system is of fundamental importance from the point of view of quantum computation and various other aspects. In practical scenario, quantum systems are sensitive to environmental interactions, which leads to the destruction of coherence. According to the standard theory of quantum measurement, each measurement results in a projection of the state vector to a particular eigenstate corresponding to the eigenvalue of the observable which has been measured. In this very process, the phase relations between different states are destroyed. In other words, the non-diagonal elements of the density matrix, which represents the quantum mechanical coherence, are destroyed. This process is known as the process of ``Decoherence" \cite{introc2}. From the point of view of both theoretical and experimental research, the methods of developing decoherence free subspaces to minimize the destruction of coherent dynamics is extremely important. For example, the spin echo, multiple pulse techniques in NMR \cite{4c1,4c2} and the method of dynamical decoupling for open quantum systems \cite{4c3,4c4} are some of the examples of various techniques for controlling decoherence. Zeno dynamics \cite{introc3,introc8} can play a very important role in controlling decoherence. Quantum Zeno effect is the inhibition of transition for decaying states due to the process of frequent observations. The short-time behavior of non-decay probability of unstable particle is shown to be non-exponential but quadratic \cite{introc15}. Wilkinson et al. \cite{4c5} observed this deviation from the usual quantum mechanical process for decay of unstable states. Misra and Sudarshan \cite{introc3} showed that this behavior, if combined with the quantum theory of measurement, will lead us to the surprising conclusion of freezing of decay dynamics due to frequent non-selective measurements. Here we intend to discuss the process of coherence control by repetitive measurement and infer the role of Zeno dynamics in sustaining the quantum coherence of the system. We consider a two level atom tunneling through a squeezed vacuum of electromagnetic field. External electromagnetic fields play the role of reservoir in open quantum systems. It is possible for squeezed vacuums to assume the role of the reservoir \cite{4c6,4c7,4c8}, though the properties of such kind of reservoirs are very much different. It is quite well known that, squeezed vacuum can have considerable effects on quantum dissipative processes \cite{4c9}, particularly, when a two-state atomic system interacts with a broadband squeezed vacuum, the transverse polarization quadratures exhibit decay processes, though different from the usual quantum decays \cite{4c6}. Squeezed vacuums are certain kind of reservoirs having strong correlations between the field amplitudes at frequencies placed symmetrically with respect to certain carrier frequency. Such an unusual reservoir exhibits a number of new features. Most of the studies dealing with the problem of a two level atom in a squeezed vacuum assume that the squeezed vacuum is broadband; i.e. the bandwidth of the vacuum is much larger than the atomic line-width and the Rabi frequency of the driving field. However, experimental realization on squeezed state indicates that the bandwidth of the squeezed light is typically of the order of the atomic line-width \cite{4c9a,4c9b,4c9c}. The most popular schemes for generating squeezed light are those using parametric oscillator operating below threshold \cite{4c9d,4c9e}, the output of which is a squeezed beam with a bandwidth of the order of the cavity bandwidth. If the oscillator works in a degenerate regime, the squeezed field has the profile with the maximum of squeezing at the central frequency and a small squeezing far from the center. In the non-degenarate regime the profile has two peaks at frequencies symmetrically displaced from the central frequency. For strong driving field and finite bandwidth of squeezing this means that the Rabi sidebands can feel very different squeezing than the central line. A realistic description of radiative properties of the two level atom in such a squeezed field must take into account the finite bandwidth of the squeezed field. Our aim is to investigate the decay dynamics of the system based on the framework of weak measurement \cite{3c3,3c4,3c5}. We have mentioned earlier that the weak value of a certain observable is the statistical average of a standard measurement procedure performed on a pre-selected and post selected (PPS) ensemble of quantum systems, given that the interaction between the measurement apparatus (or the interacting field) and each system is sufficiently weak. In our case, we interpret measurement as the interaction between the two level atomic system and the squeezed electromagnetic field. Unlike standard measurement of an observable, which sufficiently disturbs the measurement system, a weak measurement of a quantum mechanical observable done on a PPS system does not appreciably disturb the system and yields the weak value as the measured value of the observable. Since each interaction of the field or apparatus with the system is weak, any single measurement does not contain sufficient information about the system. Here an ensemble average over numerous such interactions can give us some meaningful results. So in our case, we consider numerous interactions of the squeezed field and the two level atom to infer the decay dynamics. By decay dynamics we mean both the decay of pre-selected state and the coherence destruction. In order to do that, we consider the time dependent weak value of of certain operator as described by Davies \cite{3c7}. Based on the formalism of Davies \cite{3c7}, where a generalized decay law for metastable states has been derived, we calculate the timescales for both Zeno dynamics and coherence destruction. By comparing those timescales, we derive certain condition for sustainable quantum coherence. In the next section, we discuss the master equation for the two level atom tunneling through a squeezed vacuum and find the decay parameters related to state decay and decoherence. After that we derive the weak values of the mentioned timescales and compare them to find the condition for sustainable coherence. Then we conclude with some possible implications.

\section{Master equation for two level atom tunneling through a squeezed vacuum}

Let us now consider the master equation for the two level atom tunneling through a squeezed vacuum and calculate the decay constants in terms of system and bath parameters. A very important fundamental property of squeezed states is that they reduces quantum fluctuations. The squeezing effect of electromagnetic fields has been studied by many researchers in the field of quantum optics \cite{4c7,4c10,4c11,4c6} over the past years. Here we follow the work of Tana\'{s} \cite{4c11a,4c12}, where the problem of two state system tunneling through a squeezed vacuum is dealt in the master equation approach. The Hamiltonian of the system and reservoir is defined as
\beq\label{1}
H_T=H_s+H_{r}+H_{l}+H_{int}
\eeq
where $H_s$ is the system Hamiltonian,$H_{r}$ is that of the reservoir vacuum field, $H_{l}$ is the same for the interaction between the atom and the classical laser field and $H_{int}$ is for the interaction between atom and the vacuum field respectively. They are described as
\beq\label{2}
\begin{array}{ll}
	H_s=\frac{1}{2}\hbar\omega_A\sigma_z=-\frac{1}{2}\hbar\Delta\sigma_z+\frac{1}{2}\omega_L\sigma_z\\
	H_{r}=\hbar\int_0^{\infty} d\omega \omega b^{\dag}(\omega)b(\omega)\\
	H_{l}=\frac{1}{2}\hbar\Omega[\sigma_{+}\exp(-i\omega_L t)+\sigma_{-}\exp(i\omega_L t)]\\
	H_{int}=i\hbar\int_0^{\infty} K(\omega)[b^{\dag}(\omega)\sigma^{-}-b(\omega)\sigma^{+}]d\omega
\end{array}
\eeq
where $b^{\dag}(\omega)$ and $b(\omega)$ are the creation and annihilation operators for the field, which is assumed as a collection of harmonic oscillators. In equation \ref{2}, $K(\omega)$ is the coupling of the atom to the vacuum modes and $\Delta=\omega_L-\omega_A$ is the detuning of the driving laser field frequency $\omega_L$ from the atomic frequency $\omega_A$. The laser driving field strength is given by the Rabi frequency $\Omega$. 
It is assumed that the reservoir is in a squeezed vacuum state in which the creation and annihilation operators obey the relations
\beq\label{3}
\begin{array}{ll}
	\langle b(\omega)b^{\dag}(\omega')\rangle=(N(\omega)+1)\delta(\omega-\omega')\\
	\langle b^{\dag}(\omega)b(\omega')\rangle=N(\omega)\delta(\omega-\omega')\\
	
	\langle b(\omega)b(\omega')\rangle=M\delta(2\omega_L-\omega-\omega')
\end{array}
\eeq
The functions $N(\omega)$ and $|M(\omega)|$ are the parameters describing the squeezing, which are given by
\beq\label{4}
\begin{array}{ll}
	N(\omega)=\frac{\lambda^2-\mu^2}{4}\left[\frac{1}{x^2+\mu^2}-\frac{1}{x^2+\lambda^2}\right]\\
	|M(\omega)|=\frac{\lambda^2-\mu^2}{4}\left[\frac{1}{x^2+\mu^2}+\frac{1}{x^2+\lambda^2}\right]
\end{array}
\eeq
$x=\omega-\omega_L$ is the shift from the original laser frequency $\omega_L$. $\lambda$ and $\mu$ are functions in relation to the cavity damping rate $\gamma$ and amplification coefficient $\epsilon$.
\beq\label{5}
\lambda=\gamma +\epsilon ,~~~~ \mu=\gamma -\epsilon
\eeq
and $M=|M|e^{i\phi}$. $\phi$ is considered as the phase of squeezing. The cavity dissipation constant $(\gamma)$ is related to the coupling constant $(K(\omega))$ by $\gamma=2\pi K(\Omega)^2$ \cite{4c11}. The resulting master equation is formulated as \cite{4c12}
\beq\label{6}
\begin{array}{ll}
	\dot{\rho}=\frac{1}{2}i\gamma\delta[\sigma_z,\rho]\\
	+\frac{1}{2}\gamma\widetilde{N}(2\sigma_{+}\rho\sigma_{-}-\sigma_{-}\sigma_{+}\rho-\rho\sigma_{-}\sigma_{+})\\
	+\frac{1}{2}\gamma(\widetilde{N}+1) (2\sigma_{-}\rho\sigma_{+} -\sigma_{+}\sigma_{-}\rho-\rho\sigma_{+}\sigma_{-})\\
	-\gamma\widetilde{M}\sigma_{+}\rho\sigma_{+}-\gamma\widetilde{M}^{*}\sigma_{-}\rho\sigma_{-}-\frac{1}{2}i\Omega[\sigma_{+}+\sigma_{-},\rho]\\
	+\frac{1}{4}i(\beta[\sigma_{+},[\sigma_z,\rho]]-\beta^{*}[\sigma_{-},[\sigma_{z},\rho]])
	
\end{array}
\eeq
where
\beq\label{7}
\widetilde{N}=N(\omega_L+\Omega')+\frac{1}{2}(1-\widetilde{\Delta}^2)Re\Upsilon_{-}
\eeq

\beq\label{8}
\widetilde{M}=M(\omega_L+\Omega')-\frac{1}{2}(1-\widetilde{\Delta}^2)\Upsilon_{-}+i\widetilde{\Delta}\delta_Me^{i\phi}
\eeq

\beq\label{9}
\begin{array}{ll}
	\Upsilon=N(\omega_L)-N(\omega_L+\Omega')\\
	~~~~~~-[|M(\omega_L)|-|M(\omega_L+\Omega')|]e^{i\phi}
\end{array}
\eeq

\beq\label{10}
\delta=\frac{\Delta}{\gamma}-\frac{1}{2}(1-\widetilde{\Delta}^2)Im\Upsilon_{-}+\widetilde{\Delta}\delta_N
\eeq

\beq\label{11}
\beta=\gamma\widetilde{\Omega}\left[\delta_N+\delta_M e^{i\phi}-i\widetilde{\Delta}\Upsilon_{-}\right]
\eeq

\beq\label{12}
\Omega'=\sqrt{\Omega^2+\Delta^2},~~~~\widetilde{\Omega}=\frac{\Omega}{\Omega'},~~~~\widetilde{\Delta}=\frac{\Delta}{\Omega'}
\eeq

$\delta_N$ and $\delta_M$ are the shifts induced due to squeezing. $M=|M|e^{i\phi}$, where $\phi$ is the squeezing angle.
From the master equation given by \ref{6} \cite{4c12} we can get
\beq\label{13}
\begin{array}{ll}
	\langle\dot{\sigma}_{-}\rangle=-\gamma(\frac{1}{2}+\widetilde{N}-i\delta)\langle\sigma_{-}\rangle
	-\gamma\widetilde{M}\langle\sigma_{+}\rangle+\frac{i}{2}\Omega\langle\sigma_z\rangle\\
	\langle\dot{\sigma}_z\rangle=i(\Omega+\beta^{*})\langle\sigma_{-}\rangle
	-i(\Omega+\beta)\langle\sigma_{+}\rangle\\
	~~~~~~~~~~~~-\gamma(1+2\widetilde{N})\langle\sigma_z\rangle-\gamma
	
\end{array}
\eeq
Equation for $\langle\dot{\sigma}_{+}\rangle$ is the hermitian conjugate of the equation for $\langle\dot{\sigma}_{-}\rangle$. So from \ref{13} we get
\beq\label{14}
\begin{array}{ll}
	\frac{d}{dt}(\langle\sigma_{+} + \sigma_{-}\rangle)=-\gamma(\frac{1}{2}+\widetilde{N}+\widetilde{M}^{*}-i\delta)\langle\sigma_{-}\rangle\\
	~~~~~~~~~~~~~~~~~~~-\gamma(\frac{1}{2}+\widetilde{N}+\widetilde{M}+i\delta)\langle{\sigma}_{+}\rangle\\
	~~~~~~~~~~~~~~~~~~=-\gamma\left(\frac{1}{2}+\widetilde{N}+Re\widetilde{M}\right)(\langle\sigma_{+}\rangle+\langle\sigma_{-}\rangle)\\
	~~~~~~~~~~~~~~~~~~~+i(Im\widetilde{M}+\delta)(\langle\sigma_{+}\rangle-\langle\sigma_{-}\rangle)
\end{array}
\eeq
From \ref{14} we can get the decay rate of $\langle \sigma_{+}+\sigma_{-}\rangle$ as
\beq\label{15}
\Gamma_{dec}=\gamma\left(\frac{1}{2}+\widetilde{N}+Re\widetilde{M}\right)
\eeq
We interpret this decay parameter as the decay constant associated to the destruction of coherence. $\sigma_{\pm}$ operators represent the switching of both the states from one to another. This is only possible if the system is in superposition of both the available states. Decay of the ensemble average of $\sigma_{\pm}$ means the decay of quantum superposition; i.e. the destruction of coherence.  Similarly from \ref{13} we get the decay constant for pre-selected $\sigma_z$ state as
\beq\label{16}
\Gamma_{pop}=\gamma\left(1+2\widetilde{N}\right)
\eeq
This parameter represents the population inversion from the initial pre-selected state to the other. We use these parameters to get the weak value of decohernce and zeno timescales in the next section.\section{Weak value of Decohernce time and Zeno time}
Now we are going to calculate the timescales associated to the processes of coherence destruction (decoherence) and freezing of state decay by frequent non-selective measurements (zeno effect) in the weak measurement framework. The reason behind using this particular framework is that in our consideration the interaction between the squeezed electromagnetic field and the system is sufficiently weak. So one single measurement or interaction on the system does not give any significant result; or in other words does not disturb the system in a considerable way. So numerous such interactions are taken into account to get an ensemble average for some quantum observable. Similar to what was done in the previous chapter, we use the framework originally developed by Davies \cite{3c7} to find the weak value for the decay law of metastable states. \\
Consider the time evolution for the state of the system
\beq\label{3.1}
|\psi(t)\rangle= U(t-t_0)|\psi(t_0)\rangle
\eeq
where the time evolution operator is given by
\beq\label{3.2}
U(t-t_0)=e^{-iH(t-t_0)}
\eeq
The time dependent weak value of some operator $A$ pre -selected at time $t_i$ and post selected at $t_f$ is expressed as
\beq\label{3.3}
A_w= \frac{\langle\psi_f|U^{\dagger}(t-t_f)AU(t-t_i)|\psi_i\rangle}{\langle\psi_f|U^{\dagger}(t-t_f)U(t-t_i)|\psi_i\rangle}
\eeq
Considering a two level atom in an external magnetic field $\textbf{B}$, we get the Hamiltonian of the system as
\beq\label{3.4}
H_s=-\mathbf{\mu}.\textbf{B}
\eeq
where
\beq\label{3.5}
\mathbf{\mu}=-\frac{e\hbar \textbf{S}}{2m}
\eeq
and
\beq\label{3.6}
\textbf{S}=(\sigma_x,\sigma_y,\sigma_z)
\eeq
$\sigma_{i=x,y,z}$ are the usual Pauli spin matrices.
If we consider that the magnetic field is in the z-direction, the system Hamiltonian reduces to
\beq\label{3.7}
H_s=\frac{1}{2}\hbar \omega_A \sigma_z
\eeq
The time evolution operator for the system
\beq\label{3.8}
U(t)= \left (  \begin{array}{ll}
	e^{i\omega_A t/2} & 0\\
	0 & e^{-i\omega_A t/2}
\end{array} \right)
\eeq
Let us now consider that at initial time $t_i$ the particle is x-polarized. Then we get
\beq\label{3.9a}
|\psi_i\rangle= \frac{1}{\sqrt{2}}\left (\begin{array}{ll}
	1\\
	1
\end{array} \right)
\eeq
The associated projection operator is
\beq\label{3.9b}
P_{+}= \frac{1}{\sqrt{2}}\left (  \begin{array}{ll}
	1 & 1\\
	1 & 1
\end{array} \right)
\eeq
In case of the decay of metastable states, let us consider the system is coupled to a bath of $2R$ number of environmental modes initially in their ground states. In presence of the interaction with the environmental bath modes, the system is viable to loose energy to the bath modes. Let us consider for simplicity that any arbitrary excited state $E_r$ satisfies the relation
\beq\label{3.10}
E_r-E_0=n\Delta E,~~~~~-R\leq r\leq R
\eeq
It is assumed that the reference atom is equally coupled to all the bath modes and the energy states are equispaced.
If $a_0$ is the amplitudes of the initial pre-selected state, then following Davies \cite{3c7}, we find
\beq\label{3.13}
a_0(t)= e^{-\Gamma (t-t_i)}
\eeq
where $\Gamma$ is the decay parameter. The time evolution operator $U(t)$ is a $(2N+1)\times(2N+1)$ dimensional matrix, whose components are $U_{ij}$. Now following Davies \cite{3c7}, we get
\beq\label{3.14}
U_{00}=e^{-\Gamma t}
\eeq
within the limit $\Delta E\rightarrow 0$. Under the relation $U^{\dagger}(t)=U(-t)$ the time dependent weak value of some operator $A$
\beq\label{3.16}
A_w=\frac{\langle \psi_f| U(t_f-t)AU(t-t_i)|\psi_i\rangle}{\langle \psi_f|U(t_f-t_i)|\psi_i\rangle}
\eeq
Now for our purpose of getting the weak value of survival probability, we take the operator $A$ as the projection operator $P_{+}$. Let us assume that the post selected final state at $t_f$ be $|\psi_k\rangle$.
So the weak value of the projection operator gives
\beq\label{3.18}
P_w=\frac{U_{k0}(t_f-t)U_{00}(t-t_i)}{U_{k0}(t_f-t_i)}
\eeq
Now we want to get expression for the weak value of the survival probability for the pre-selected state. So choosing $E_k=E_0$ (ie. the post selected state is the same as the pre selected state), we get the simple expression
\beq\label{3.20}
P_w=e^{-\Gamma(t-t_i)} \left[\frac{1-e^{-\Gamma(t_f-t)}}{1-e^{-\Gamma(t_f-t_i)}}\right]
\eeq
The total time interval which consists of $n$ successive measurements is taken as $t_f-t_i=n\tau_M$. Where $\tau_M$ is the time interval between two successive measurements. This time interval is nothing but the interval between two successive interaction with the field. From \ref{3.20} we get
\beq\label{3.20a}
\begin{array}{ll}
	P_w=1~~~~~ for~~t=t_i\\
	~~~~=0~~~~for~~t=t_f
\end{array}
\eeq
So \ref{3.20} gives the generalized decay law in weak measurement scheme. At the instant of initial measurement ($t_i$), the particle is in the pre selected state and at the instant of final measurement ($t_f$), the particle is decayed due to the interaction with the environment. So we get the decay time by integrating over the weak survival probability within the limit $t_i$ to $t_f$
\beq\label{3.21}
\tau_{decay}=\int_{t_i}^{t_f} e^{-\Gamma(t-t_i)} \left[\frac{1-e^{-\Gamma(t_f-t)}}{1-e^{-\Gamma(t_f-t_i)}}\right]dt
\eeq
The amplitude of the squeezed field varies with it's frequency $\omega_L$. The amplitude of the field becomes maximum after this interval. So this time period equaling to the inverse of $\omega_L$ can also be interpreted as the time period of maximum interaction. Therefore we infer that the measurement time is equal to the inverse of $\omega_L$ .Again here we are considering the Zeno dynamics; i.e. frequent observation. So the measurement time is considered to be quite small. Under the assumption $\tau_M\ll1/\Gamma$, from \ref{3.21} we get the decay time
\beq\label{3.22}
\tau_{decay}=\frac{1}{\Gamma+2\omega_L/n}
\eeq
Here we have taken $\tau_M=1/\omega_L$. Now putting the coherence decay parameter $\Gamma_{dec}$ from \ref{15} of the previous section in \ref{3.22}, we get the weak value of decoherence time as
\beq\label{3.23}
\tau_{dec}=\frac{1}{\gamma\left(\frac{1}{2}+\widetilde{N}+Re\widetilde{M}\right)+2\omega_L/n}
\eeq
This is the timescale within which the system loses it's coherence; i.e. the non-diagonal components of the reduced density matrix of the system vanishes. \\
Now we turn our attention to calculate the Zeno time for the concerning system. As we have mentioned earlier, the quantum zeno effect is the inhibition of transition of metastable states under frequent observation, so the timescale for Zeno effect is the timescale within which the system stays in it's initial state. We get the weak value of this particular timescale by putting the expression of state decay parameter from \ref{16} in \ref{3.22} as
\beq\label{3.24}
\tau_{zeno}=\frac{1}{\gamma\left(1+2\widetilde{N}\right)+2\omega_L/n}
\eeq
In our understanding, the dynamics of decoherence and Zeno effect has got an intrinsic reciprocal relation. Let us elaborate the reason behind this statement. Whenever any kind of disturbance in the form of interaction dominates the time evolution of the state, the concerning system evolves in a reduced subspace of the total Hilbert space \cite{introc8}. This reduced subspace is known as ``Zeno subspace". The appearance of these kind of subspaces is caused by frequent non-selective measurement. Frequent measurements splits the total Hilbert space into these subspaces, within which leakage of probability is not possible. So it can be argued that each of these reduced subspaces acts like an isolated space, within which environmental interaction is absent, or at least minimum. Under very strong environmental interaction, these subspaces cannot be sustained due to extreme decoherence. It is our inference that the timescale characterizing Zeno effect must give a certain lower limit to decoherence, below which the process of decoherence is very fast and uncontrollable. So to control the process of decoherence by means of frequent non-selective measurement, we must state a precondition that the decoherence time must be greater than the zeno time for the system.\\
Comparing \ref{3.23} and \ref{3.24} we get
\beq\label{3.25}
\frac{\tau_{zeno}}{\tau_{dec}}=\frac{1}{2}+\frac{2\gamma Re\widetilde{M}+\omega_L/n}{\gamma(1+2\widetilde{n})+2\omega_L/n}
\eeq
As per our argument given above: $\tau_{dec}\geq \tau_{zeno}$. Under this condition sustainable coherent dynamics is possible. Imposing this condition on \ref{3.25}, we get
\beq\label{3.26}
4Re\widetilde{M}\leq (1+2\widetilde{N})
\eeq
From \ref{3.26} we can calculate further to get
\beq\label{3.27}
\frac{\sqrt{\widetilde{\Delta}^2\delta_M^2+|M(\omega_L+\Omega')|^2}\sin(\theta-\phi)}{1+2N(\omega_L+\Omega')+3(1-\widetilde{\Delta}^2)Re\Upsilon_{-}}\leq\frac{1}{4}
\eeq
where
\beq\label{3.27a}
\tan\theta=\frac{|M(\omega_L+\Omega')|}{\widetilde{\Delta}\delta_M}
\eeq
A special case of squeezing phase \cite{4c13} can be set as
\beq\label{3.28}
\phi(\Delta)=\frac{\pi\Delta}{\Omega}
\eeq
The condition \ref{3.27} will surely hold if $\theta\rightarrow\phi$.
In our consideration, the cavity damping constant ($\gamma$) and the real amplification constant ($\epsilon$) are small compared to the Rabi frequency ($\Omega$). So $\Omega\gg \mu,\lambda$. Under this condition, using the expressions of $\delta_M$ from \cite{4c12}, we find that
\beq\label{3.29}
\tan\theta= \frac{1}{\Omega'\widetilde{\Delta}}\frac{\mu\lambda}{\mu+\lambda}=\frac{\gamma^2-\epsilon^2}{2\Delta\gamma}
\eeq
The sufficient condition for Zeno effect to dominate over decoherence can be stated as
\beq\label{3.30}
\frac{\gamma^2-\epsilon^2}{2\gamma}\rightarrow\Delta\tan\left(\frac{\pi\Delta}{\Omega}\right)
\eeq
For a special case, if $\Delta\ll\Omega$, we find that the condition will hold if $\gamma\sim\epsilon$. i.e. for the case where the shift of laser and atomic frequency is very small compared to the Rabi frequency and for small dissipation, the Zeno dynamics will dominate over the decoherence dynamics, if the real amplification constant is comparable in magnitude to the cavity dissipation parameter. Under this condition, sustainable coherent dynamics can be achieved.

\section{Conclusion}

In this work, we have done a comparison between decoherence and Zeno dynamics in the framework of weak measurement. The weak value of decoherence time and Zeno time has been calculated and compared. Based on the assertion that decoherence time must be longer than Zeno time for the Zeno effect to dominate over decoherence, we found a certain condition under which we have sustainable coherent dynamics. Here we have considered the system of a two level atom tunneling through a squeezed vacuum. These type of systems are used to develop optical ion traps, which are important components for developing quantum memory devices. But the challenge for building quantum memory devices is that of controlling decoherence effect to sustain the ``quantumness" of the system. In this work, we have shown that Zeno dynamics can be an effective process to control such environmental effects. The reason behind using the weak measurement framework is also necessary  to mention here. In this type of measurement scheme, the interaction between the system and the measuring device (in our case the squeezed field) is made very small. This is a good way to minimize the environmental interaction, which is in turn helpful for reducing the effect of decoherence. Another important feature of weak measurement process is that, here we take an ensemble average of numerous observations over the pre-selected and post-selected states, because one single measurement interaction cannot bring out enough information about the system. Since the Zeno dynamics is initiated by frequent observations, an ensemble average over many such observations is necessary to observe the dynamics over a finite period of time. So weak measurement scheme is also very much compatible with the Zeno type measurement procedure. Now we have mentioned earlier that controlling decoherence dynamics is essential to build quantum memory devices. As we have shown in this work that this type of effect can be minimized using certain procedures like frequent non-selective measurement and weak measurement scheme, but effectively it cannot be eradicated completely. So it is important to construct the quantum computational schemes in the backdrop of open quantum systems. Considering these facts, we intend to extend our formalism to the area of adiabatic quantum computation in future publications.

\section{Acknowledgement}

The author thanks Prof. Sisir Roy of Indian Statistical Institute for helpful discussions.

\end{document}